# Nickel Foam as a Substrate for III-nitride Nanowire Growth

Michael A. Mastro[1*], Neeraj Nepal[1], Fritz Kub[1], Jennifer K. Hite[1], J. Kim[2], and Charles R. Eddy, Jr.[1]

[1]U.S. Naval Research Laboratory, 4555 Overlook Ave., SW, Washington, D.C. 20375

[2]Department of Chemical and Biological Engineering, Korea University, Seoul, South Korea

**Abstract**

This article presents the use of flexible metal foam substrates for the growth of III-nitride nanowire light emitters to tackle the inherent limitations of thin-film light emitting diodes as well as fabrication and application issues of traditional substrates. A dense packing of gallium nitride nanowires were grown on a nickel foam substrate. The nanowires grew predominantly along the a-plane direction, normal to the local surface of the nickel foam. Strong luminescence was observed from undoped GaN and InGaN quantum well light emitting diode nanowires.

**Introduction**

The GaN-based light emitting diode (LED) market has grown into a multi-billion dollar market in just the last two decades. Despite this rapid progress, certain restrictions are inherent to the thin-film on a planar substrate design. These constraints can be generalized into light extraction, defectivity, substrate cost, and processing cost limitations as well as a lack of mechanical flexibility.[1]

Sapphire is the predominant substrate for epitaxy of III-nitride light emitting diode thin films. Sapphire is non-conductive and presents a large lattice mismatch with the III-nitride material system. Silicon carbide substrates are expensive but have a closer lattice mismatch to GaN and can be supplied in a conductive state.[2-4] Silicon is available in larger diameters but GaN epitaxy on silicon suffers from thermal stress constraints.[5] The cost of the wafer is significant but often overstated when compared to the processing and balance of system costs.[6] A move to larger wafers allows a significant increase in back-end processing throughput and commensurate decrease in the cost per die.[7]

An ongoing idea is to produce III-nitride LEDs on low-cost, large-area substrates such as glass akin to the thin film photovoltaic technologies.[8,9] Despite some progress for producing III-nitride LEDs on glass,[10] a poly-crystal type growth is inherently produced when the underlying substrate, such as glass, does not present crystalline order to which reactant atoms can align to form an ordered thin film. Poly-crystal or fine-grain III-nitride material presents an exceedingly large number of dislocations and other defects that effectively destroy operation of the pn junction.[11]

Even for high-quality thin film growth on sapphire, lattice mismatch leads to the formation of dislocations with densities greater than $10^8$ cm$^{-2}$, which limit the internal quantum efficiency of LEDs.[12] Furthermore, green LEDs require high indium content in the $In_xGa_{1-x}N$ quantum wells that under planar lattice stress can encourage the formation of V-pits. Moreover, $In_xGa_{1-x}N$ stability is reduced at the elevated growth temperature needed for thin film metal organic chemical vapor deposition (MOCVD).

Another intrinsic issue with a standard LED structure is the low external extraction efficiency of light owing to the index of contrast difference with air.[13] A majority of the light generated in a semiconductor thin film on a planar substrate suffers from total internal reflection. Complex procedures such as die shaping, photonic crystals, micro-cavities, and surface roughening are used to extract light that would normally be trapped in the semiconductor and substrate slab.[14-19]

Nanowire III-nitride light emitters have been suggested and demonstrated to avoid many of the deleterious issues associated with thin film structures.[20,21] The dimensions of the nanowire are on the order of the optical wavelength; therefore, light is easily scattered out of the semiconductor into the surrounding air. The growth of nanowires via vapor-liquid-solid mechanism proceeds at temperature lower than that used for thin-film MOCVD easing the incorporation of indium into the active region.[22] The removal of the in-plane lattice constraint allows nanowires to grow with a greatly reduced defect level relative to its thin-film equivalent. It is known that the vapor-liquid-solid growth of GaN nanowires is known to have a lesser dependence on the underlying substrate.[23]

Recently, a two-step reactive vapor / MOCVD approach was used to grow GaN nanowires with an InGaN shell on a stainless steel substrate.[24] In general, a metal substrate can be scaled to any reasonable process tool size or shape.

This article presents a nanowire LED design based on a metal foam substrate. In contrast to a solid metal substrate, the foam form further lowers the amount and cost of material used for a given surface area. Furthermore, the suppleness of the foam substrate extends the application space to other areas including coiled piezoelectric energy harvesting devices and flexible light sources. It is expected from simple ray tracing that the vast majority of light generated from the nanowires will not be reabsorbed somewhere else in the architecture. This includes light generated deep in the structure that can be expected to directly escape through the micron-scale pores of the foam substrate.

**Experimental**

A 0.05 M nickel nitrate solution was repeatedly dripped onto a nickel foam substrate and blown dry in $N_2$ then loaded into a vertical impinging-flow MOCVD reactor. A 50-Torr, $N_2/H_2$ mixed atmosphere was used during the ramp to growth temperature. Trimethylgallium was flowed for 2 sec prior to the onset of $NH_3$ flow to prevent nitridation of the nickel seeds. The GaN nanowire core was grown at a temperature of 850°C, a pressure of 50 Torr and a V/III ratio of 50.[25] Under proper growth conditions, the metal catalyst particle captures reactants and enhances the growth rate perpendicular to the substrate, thus creating a pseudo-one-dimensional semiconductor wire. The LED nanowires continued with growth of an InGaN quantum well shell at 600°C and a V/III ratio of 150. Immediately thereafter, the GaN:Mg shell was grown to avoid decomposition of the InGaN. The introduction of Mg dopant atoms is known to encourage a higher lateral growth rate relative to growth of undoped GaN nanowires under equivalent conditions.[26,27] The samples were cooled from growth temperature in pure nitrogen ambient to avoid rapid decomposition in hydrogen and, in the case of the samples with a p-type shell, to activate the acceptor dopant (Mg). Structural characterization was performed with a LEO FE Scanning Electron Microscope (SEM). Photoluminescence (PL) measurements were carried out using a HeCd laser at 325 nm and an Ocean Optic QE6500 spectrometer.

**Results and Discussion**

Images of the GaN nanowires on the nickel foam framework are observable in a series of scanning electron micrographs in Fig 1. The nanowires grow uniformly over the entire foam surface. This particular foam has pores with an average diameter of 200 μm although a similar uniform coating was achieved for pore diameters down to 5 μm. At smaller pore volumes, it is possible to form a coalesced film.

The direction of GaN growth is generally perpendicular to the local surface. Growth of III-nitride nanowires via a VLS mechanism under this growth condition tends to proceed in the <11-20> a-direction with an isosceles triangular cross-section exhibiting a distinct set of facets of (0001), (1-10-1), and (-110-1).

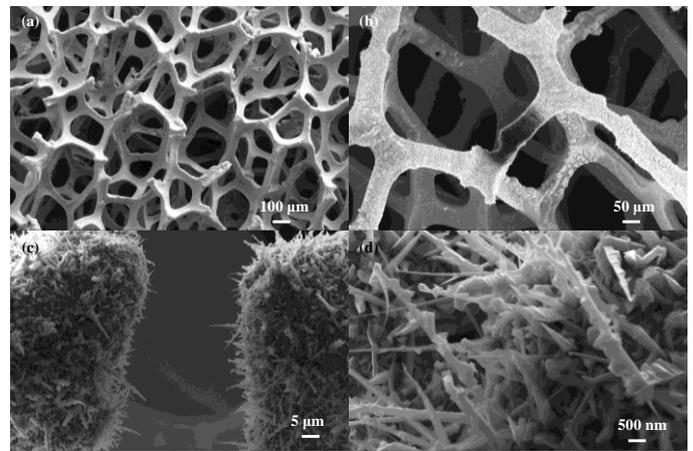

Figure 1. Electron micrograph of GaN nanowires on a nickel foam substrate at increasing level of magnification.

The GaN wire length is proportional to growth time with 1 hr of growth yielding wires of approximately 10 μm in length. The GaN nanowires have an approximate 200 nm diameter over a length of several microns. A slight tapering is evident owing to growth at an elevated nanowire growth temperature although these conditions are known to produce higher quality GaN nanowires.[28]

While the nanowires grow along the a-direction, the curvature of the underlying foam presents a range of diffracting planes at the nominal surface. A 2θ-θ X-ray diffraction pattern in Fig. 2 presents sharp diffraction from the (10-10), (0002), and (10-10) planes as well as weaker diffraction from higher index planes.

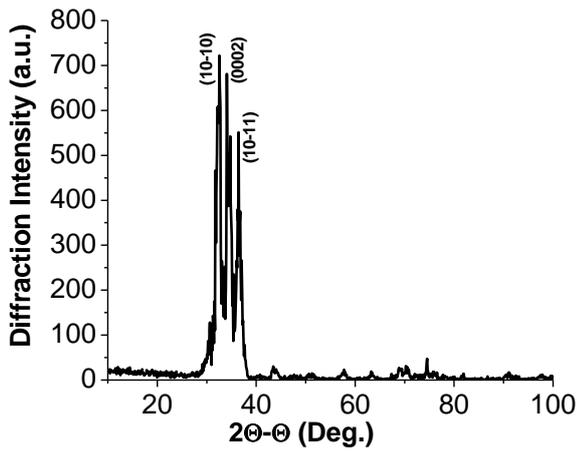

Figure 2. X-ray diffraction pattern of GaN nanowires on a nickel foam substrate. The nickel foam form intrinsically possesses a continuum of surface normals. Thus, nanowires grown on this curved surface present a spectrum of GaN crystal planes. Low index GaN diffraction planes dominate the pattern although higher index planes are also evident.

The room temperature PL of the GaN nanowires in Fig. 3 displays two major bands in the spectrum corresponding to the band-edge and the donor-acceptor luminescence transitions. The dominant band at 3.4 eV is associated with recombination processes involving the annihilation of free-excitons and a band-to-band transition. The enhancement of the higher energy side of the band-edge PL emission is typically attributed to the high-excess of free electron carriers in high-quality GaN.[21]

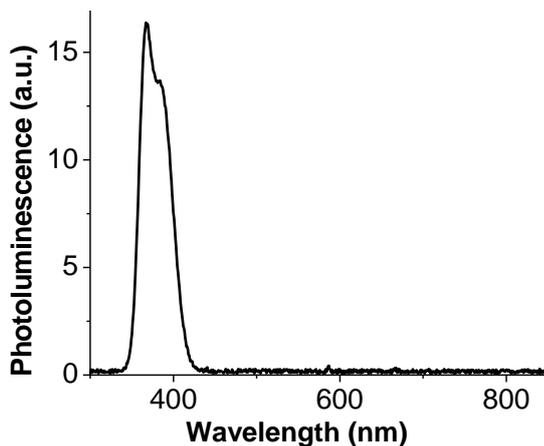

Figure 3. Photoluminescence intensity of GaN nanowires on a nickel foam substrate. The near band-edge emission dominates the spectrum.

Similar samples as described above were fabricated with the addition of n-type doping to the GaN:Si core followed by the formation of an InGaN-well/GaN:Mg shell around the core. The thickness of the InGaN shell layer is approximately 5 nm and, the outer thickess of the GaN:Mg layer is 200 nm after 5 min of growth. The thickness of the InGaN/GaN:Mg sheath was directly proportional to growth time. The structure of the nano-wires was designed to create a thin InGaN well for quantum confinement of the injected carriers, and a thicker GaN:Si core and GaN:Mg sheath for optical confinement of the optical mode.

Fig. 4 shows the PL spectrum from GaN:Mg / InGaN well (sheath) / GaN:Si (core) nano-wires. The nano-wires display the expected GaN band-edge (near 3.4 eV) and near band-edge dopant-based transitions as well as the 542 nm InGaN luminescence from the quantum well. Likely, the emission of the InGaN is slightly blue-shifted by the onset of quantum confinement in the 5-nm well.

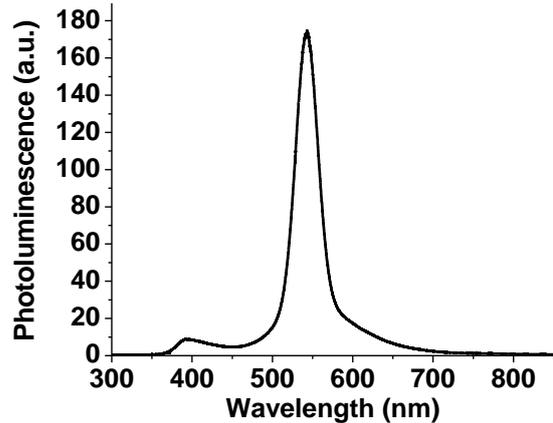

Figure 4. Photoluminescence spectrum of GaN:Mg / InGaN well / GaN:Si core-shell nanowires on a nickel foam substrate.

## Conclusion

The focus of this investigation was to demonstrate high-quality, reproducible group-III nitride nano-wire emitters on a conductive, low-cost, and flexible metal foam substrates.

## Acknowledgements

Research at the US Naval Research Lab is partially supported by the Office of Naval Research.